\documentclass[11pt, a4paper]{article}
\usepackage[breaklinks]{hyperref}  
\hypersetup{colorlinks,urlcolor=black,citecolor=black,linkcolor=black,filecolor=black}
\usepackage{breakurl}

\usepackage{amsmath,amssymb,booktabs}

\usepackage[pdftex]{graphicx,color}
\usepackage{epstopdf}
\begin{document}

\thispagestyle{plain}

\def\bib{B\kern-.05em{I}\kern-.025em{B}\kern-.08em}
\def\btex{B\kern-.05em{I}\kern-.025em{B}\kern-.08em\TeX}





\title{A Solution to Lithium Problem by Long-Lived Stau}

\author{
\centerline{
Joe Sato$^{1}$\footnote{E-mail address: joe@phy.saitama-u.ac.jp}
,~
Takashi Shimomura$^{2}$\footnote{E-mail address: shimomura@cc.miyazaki-u.ac.jp},
~and 
Masato~Yamanaka$^{3}$\footnote{E-mail address: masato.yamanaka@cc.kyoto-su.ac.jp}
}
\\*[23pt]
\centerline{
\begin{minipage}{\linewidth}
\begin{center}
$^1${\it \normalsize Department of Physics, Saitama University, \\
Shimo-okubo, Sakura-ku, Saitama, 338-8570, Japan}  
\\*[10pt]
$^2${\it \normalsize Faculty of Education, University of Miyazaki, \\
Miyazaki, 889 -2192, Japan}  
\\*[10pt]
$^3${\it \normalsize Kobayashi Maskawa Institute, Nagoya University, \\
Nagoya 464-8602, Japan, 
\\
Maskawa Institute, Kyoto Sangyo University, \\
Kyoto 603-8555, Japan}  
\end{center}
\end{minipage}}
\\*[50pt]}
\date{}
\maketitle

\begin{abstract}
We review a non-standard Big-Bang Nucleosynthesis (BBN) scenario within 
the minimal supersymmetric standard model, and propose an idea to solve 
both ${}^{7}\text{Li}$ and ${}^{6}\text{Li}$ problems. Each problem 
is a discrepancy between the predicted abundance in the standard BBN 
and observed one. 
We focus on the stau, a supersymmetric partner of tau lepton, which is a 
long-lived charged particle when it is the next lightest supersymmetric 
particle and is degenerate in mass with the lightest supersymmetric particle. 
The long-lived stau forms a bound state with a nucleus, and provide 
non-standard nuclear reactions. One of those, the internal conversion process, 
accelerates the destruction of ${}^{7}\text{Be}$ and ${}^{7}\text{Li}$, 
and leads to a solution to the ${}^{7}\text{Li}$ problem. 
On the other hand, the bound state of the stau and ${}^{4}\text{He}$ 
enhances productions of n, D, T, and ${}^{6}\text{Li}$. The 
over-production of ${}^{6}\text{Li}$ could solve the ${}^{6}\text{Li}$ 
problem. While, the over-productions of D and T could conflict with 
observations, and the relevant parameter space of the stau is strictly 
constrained. We therefore need to carefully investigate the 
stau-${}^{4}\text{He}$ bound state to find a condition of solving the 
${}^{6}\text{Li}$ problem.
The scenario of the long-lived stau simultaneously and successfully fit the 
abundances of light elements (D, T, ${}^{3}\text{He}$, ${}^{4}\text{He}$, 
${}^{6}\text{Li}$, and ${}^{7}\text{Li}$) and the neutralino dark 
matter to the observed ones.  Consequently parameter space both of the 
stau and the neutralino is determined with excellent accuracy. 
\end{abstract}


\section{Introduction}	

Lithium problem is a long standing question\cite{Spite:1982dd},
which is a discrepancy of the Lithium primordial abundance between
observed values and the predicted value by 
the standard big-bang nucleosynthesis
(SBBN).
The observed values of $^7$Li/H are inferred from metal-poor stars
and are set as (1-2)$\times 10^{-10}$,~\cite{Melendez:2004ni, 
Bonifacio:2006au, Monaco:2010mm, Monaco:2011sd, Aoki:2012wb, 
Mucciarelli:2014} while the theoretical prediction in the SBBN is
$(5.24\pm 0.7)\times 10^{-10}$~\cite{Coc:2011az,Cyburt:2008kw}.
There is discrepancy larger than 4$\sigma$ level.
This is often called $^7$Li problem.

There is also a tension in $^6$Li primordial abundance between the 
observed and the predicted values.
The observed value is about 1000 times higher than the predicted one
\cite{Asplund:2005yt}, though at this moment it is still controversial 
\cite{Lind:2013iza}.

The predictions by the SBBN are given
using the standard model (SM) of the particle physics.
Though there is still a room for astrophysical explanation 
of these discrepancies~\cite{Richard:2004pj},  we take it
as an evidence which asks us to extend the SM.
Since the big-bang nucleosynthesis (BBN) took place from 1 to 1000 sec. after the big bang,
it means that we have to add a long-lived particle.
Indeed, for example, long-lived massive charged particles
induce non-standard nuclear reactions in the BBN, and drastically change  light element
abundances~\cite{Khlopov:1984pf, Pospelov:2006sc, Kohri:2006cn, Kaplinghat:2006qr,
Cyburt:2006uv, Steffen:2006wx,Bird:2007ge, Kawasaki:2007xb,
Hamaguchi:2007mp,
Jittoh:2007fr, Jedamzik:2007cp, Pradler:2007is, Kawasaki:2008qe,
Jittoh:2008eq, Pospelov:2008ta, Kamimura:2008fx, Kusakabe:2008kf,
Bailly:2008yy, Bailly:2009pe, Kusakabe:2010cb,
Jittoh:2011ni,Cyburt:2012kp,Kusakabe:2013tra}.

The SM has had enormous successes in describing the interactions of  
the elementary particles, predicting almost every experimental results 
with accuracy.  The recent discovery of the Higgs particle finally crowned 
the accumulation of successes~\cite{Aad:2012tfa, Chatrchyan:2012xdj}. 
However, a number of questions are still left that suggest the presence 
of a more fundamental theory behind. Among such questions is the 
nature of dark matter; it became a compelling question during this 
decade after the precise observations of the universe had reported their
results~\cite{Bennett:2012zja,Ade:2013zuv}. 
Though its identity is still unknown, one of the most prominent 
candidates is weakly interacting massive particles 
(WIMPs)~\cite{Bergstrom:2000pn, Munoz:2003gx, Bertone:2004pz, 
Feng:2010gw}. As is well known, supersymmetric (SUSY) extensions 
of the SM provide a stable exotic particle, the lightest supersymmetric 
particle (LSP), if R parity is conserved. Among LSP candidates, the 
neutralino LSP is the most suitable for non-baryonic dark matter since
its nature fits that of the WIMPs~\cite{Goldberg:1983nd,Ellis:1983ew}.
Neutralinos are a linear combination of neutral fermions which are the 
supersymmetric partners of hypercharge gauge boson (bino) , weak 
one (wino), and Higgs bosons (higgsinos).

In many scenarios the lightest neutralino LSP, $\tilde{\chi}^{0}_{1}$,  
consists almost of bino. 
In this case, the next-to-LSP (NLSP) is required to be degenerate in mass
to predict the appropriate dark matter density employing
coannihilation mechanism~\cite{Griest:1990kh}. 
Stau $\tilde{\tau}$, a scalar partner of tau lepton, is the most likely 
candidate for the NLSP~\cite{Ellis:1999mm}. Their mass difference, 
$\delta m = m_{\tilde{\tau}} - m_{\tilde{\chi}^0_1}$, must satisfy 
$\delta m/m_{\tilde{\chi}^0_1} <$ a few \%, where $m_{\tilde\tau}$ 
is the NLSP (stau) mass and $m_{\tilde{\chi}^0_1}$ is the LSP (lightest neutralino) 
mass. 
Observed dark matter abundance can be yielded even in 
$\delta m/m_{\tilde{\chi}^0_1} \simeq 0$. 
If $\delta m$ is smaller than tau mass, $m_\tau$, stau cannot decay into 
2 body and become very long-lived~\cite{Profumo:2004qt, Jittoh:2005pq} 
assuming lepton number conservation.
The lifetime can be very long so that the stau can change the SBBN 
prediction. This stau can effectively destroy $^7$Be which becomes 
$^7$Li by electron capture reaction after the BBN era. Since at the BBN era would-be $^7$Li exists 
as $^7$Be, to destruct $^7$Be effectively means to reduce $^7$Li 
primordial abundance. This long-lived stau with the degenerate mass can offer 
the solution to $^7$Li problem~\cite{Jittoh:2007fr, Jittoh:2008eq, 
Jittoh:2010wh, Jittoh:2011ni, Kohri:2012gc, Konishi:2013gda, Kohri:2014jfa}.
We review this idea in this article assuming exact tau number
conservation.

In Sec.~2 we introduce the interactions relevant with our scenario and
review why the stau is long-lived. Next in Sec.~3 we expound new
reactions to the nucleosynthesis chain in the SBBN.  
Then in Sec.~4 we show numerical results and their interpretation.
Finally in Sec.~5 we summarize the scenario and give a brief review 
when we introduce tiny lepton flavor violation. By introducing it the 
lithium problems become all clear with sacrifice of adding one parameter.

\section{Long-lived stau in stau-neutralino coannihilation scenario}  \label{Sec:long}  

In this section, first, we summarize interactions to formulate decay processes 
of the stau and non-standard nuclear reactions in the bound states. Then we 
briefly review a scenario that predicts the long-lived stau, and we calculate 
the lifetime.

\subsection{Interactions}  \label{Sec:interaction}  

Decay processes of the stau and the non-standard nuclear reactions in a 
stau-nucleus bound state are described by the following Lagrangian,
\begin{equation}
\begin{split}
	\mathcal{L}
	&
	= \tilde{\tau}^{*} \overline{ \tilde{\chi}_{1}^{0} } 
	(g_\text{L} P_\text{L} + g_\text{R} P_\text{R}) \tau   
	+ \sqrt{2} G_\text{F}
	\nu_\tau \gamma^\mu P_\text{L} \tau J_\mu    
	\\& \hspace{8mm} 
	+ \frac{4G_\text{F}}{\sqrt{2}} 
	\left( \bar{l} \gamma^{\mu} P_\text{L} \nu_{l} \right) 
	\left( \bar{\nu}_{\tau} \gamma_{\mu} P_\text{L} \tau \right) 	
	+ \text{h.c.}, 
\end{split}     \label{Lag}
\end{equation} 
where $G_{\textrm{F}} = 1.166 \times 10^{-5} \mathrm{GeV^{-2}}$ is the 
Fermi coupling constant, $P_{\text{L(R)}}$ represents the chiral projection operator, 
and $J_{\mu}$ represents the weak hadronic current. The effective coupling constants 
$g_\text{L}$ and $g_\text{R}$ are given by 
\begin{equation}
\begin{split}
    g_\text{L} 
    = \frac{g}{\sqrt{2}}  
    \tan\theta_\text{W} \cos\theta_\tau , ~   
    g_\text{R} 
    = - \sqrt{2} g \tan\theta_\text{W}  
    \sin\theta_\tau 
    \mathrm{e}^{i \gamma_\tau}, 
\end{split}
\end{equation}
where $g$ is the $SU(2)_\text{L}$ gauge coupling constant and
$\theta_{\textrm{W}}$ is the Weinberg angle. The mass eigenstate of staus 
is given by the linear combination of $\tilde \tau_{\textrm{L}}$ and $\tilde 
\tau_{\textrm{R}}$, the superpartners of left-handed and right-handed tau 
leptons, as
\begin{equation}
\begin{split}
	\tilde \tau 
	= 
	\cos\theta_\tau \tilde \tau_\text{L} 
	+ \sin\theta_\tau \mathrm{e}^{-i \gamma_\tau} \tilde \tau_\text{R} . 
\end{split}
\end{equation}
Here $\theta_\tau$ is the left-right mixing angle of staus and $\gamma_\tau$ 
is the CP violating phase.

\subsection{Long-lived stau}  \label{Sec:long}  

\begin{figure}[ht]
\begin{center}
\includegraphics[width=120mm]{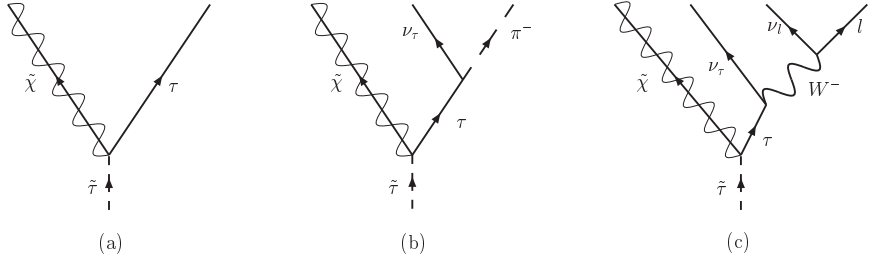}
\caption{Feynmann diagrams of stau decay: 
(a) $\tilde{\tau} \to \tilde{\chi}_{1}^{0} + \tau$, 
(b) $\tilde{\tau} \to \tilde{\chi}_{1}^{0} + \nu_{\tau} + \pi$, 
(c) $\tilde{\tau} \to \tilde{\chi}_{1}^{0} + l + \nu_{\tau} + \nu_{l}$. 
Here $l \ni \{e, \mu\}$. 
}
\label{Fig:diagrams}
\end{center}
\end{figure}

\begin{figure}[ht]
\begin{center}
\includegraphics[width=100mm]{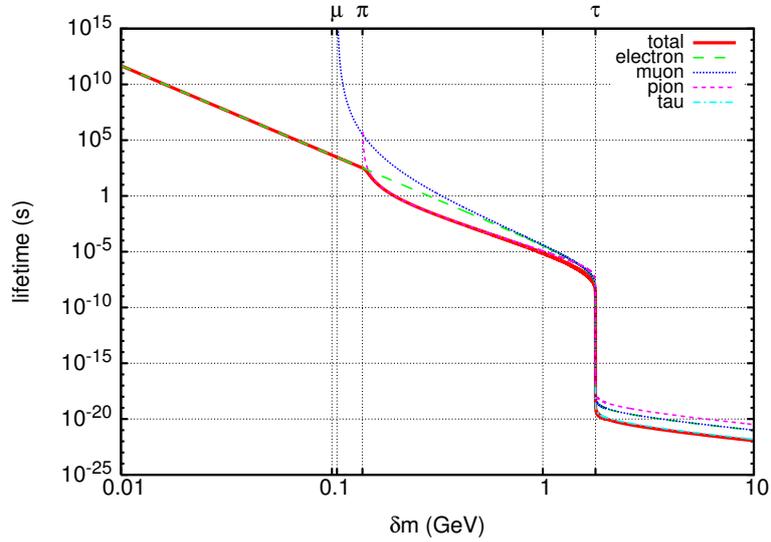}
\caption{Stau lifetime as a function of $\delta m = 
m_{\tilde{\tau}} - m_{\tilde{\chi}_{1}^{0}}$. 
We take $m_{\tilde{\tau}} = 300\,\text{GeV}$, 
$\theta _{\tau}=\pi /3$, and $\gamma_{\tau}=0$. 
Each line shows partial lifetimes of each decay channel; 
$\tilde{\tau} \to \tilde{\chi}_{1}^{0} + \tau$ (tau), 
$\tilde{\tau} \to \tilde{\chi}_{1}^{0} + \nu_{\tau} + \pi$ (pion), 
$\tilde{\tau} \to \tilde{\chi}_{1}^{0} + \nu_{\tau} + \mu + \bar{\nu}_{\mu}$ (muon), 
$\tilde{\tau} \to \tilde{\chi}_{1}^{0} + \nu_{\tau} + e + \bar{\nu}_{e}$ (electron). 
This figure is taken from Ref.~\cite{Jittoh:2005pq}. }
\label{Fig:lifetime}
\end{center}
\end{figure}

In the minimal supersymmetric standard model with R-parity conservation, the bino-like neutralino is a WIMP 
candidate. Direct search results for SUSY particles at the LHC indicate the 
neutralino with the mass of $m_{\tilde{\chi}_{1}^{0}} \gtrsim \mathcal{O} 
(100\,\text{GeV})$~\cite{Aad:2015baa}. Naive calculations 
of the relic abundance of such the heavy bino-like neutralino dark matter result in an 
over-closure of the universe. 
The coannihilation mechanism successfully  yields the observed relic abundance of the 
neutralino dark matter~\cite{Griest:1990kh, Ellis:1998kh, Edsjo:2003us}. 
The key ingredient for the mechanism is the degeneracy in mass between the 
neutralino dark matter and the NLSP stau, which requires 
$(m_{\tilde{\tau}} - m_{\tilde{\chi}_{1}^{0}})/m_{\tilde{\chi}_{1}^{0}} 
\lesssim \mathcal{O}(\%)$.  The degeneracy leads to comparable number 
densities of them at decoupling of the neutralino dark matter, and makes the 
effective annihilation rate of the dark matter large through the stau-stau and 
stau-neutralino annihilation channels. 
Such a tight degeneracy makes the stau to be long-lived by kinematical suppression 
in its decay~\cite{Profumo:2004qt, Jittoh:2005pq}. 
Especially important and interesting parameter space is wherein the mass 
difference of the stau and the neutralino, $\delta m = m_{\tilde{\tau}} 
- m_{\tilde{\chi}_{1}^{0}}$, is smaller than tau mass. 
\footnote{
Indeed in large part of parameter space wherein the Higgs mass is consistent with 
the reported one, the mass difference between the neutralino and the stau is smaller
than the mass of tau lepton~\cite{Aparicio:2012iw,Citron:2012fg,Konishi:2013gda}. 
Collider phenomenology in such parameter region is extensively 
studied~\cite{Kaneko:2008re, Heisig:2011dr, Ito:2011xs, Hagiwara:2012vz, 
Acharya:2014nyr, Desai:2014uha, Heisig:2015yla}. 
}
In such a parameter space the 2-body decay channel (diagram (a) in 
Fig.~\ref{Fig:diagrams}) closes, and open decay channels are 3- and 4-body 
ones (diagram (b) and (c) in Fig.~\ref{Fig:diagrams}, respectively).

The stau lifetime is shown by red solid line in Fig.~\ref{Fig:lifetime}. Partial 
lifetimes of each decay channel are also plotted. The lifetime is drastically enhanced 
at $\delta m = m_{\tau}$, where the 2-body decay channel closes. Partial 
widths of 3- and 4-body ones are suppressed by higher order couplings and 
tight phase space of final states. The dependence of SUSY particle masses on the partial widths 
is 
$\Gamma (\tilde{\tau} \to \tilde{\chi}_{1}^{0} \nu_{\tau} \pi) 
\propto \left( \delta m \right) \left( (\delta m)^2 - m_{\pi}^2 \right)^{5/2} 
/ m_{\tilde{\tau}}$ 
and 
$\Gamma (\tilde{\tau} \to \tilde{\chi}_{1}^{0} \nu_{\tau} l \bar{\nu}_{l}) 
\propto \left( \delta m \right)^{3} \left( (\delta m)^2 - m_{l}^2 \right)^{5/2} 
/ m_{\tilde{\tau}}$. 
Analytic calculations and more detailed analysis of the partial widths are given in 
Ref.~\cite{Jittoh:2005pq}.

The BBN took place from $\sim 1$\,second to $\sim 10$\,minutes after the big 
bang. Thermally produced stau still exists in the BBN epoch in the parameter region 
of $\delta m \lesssim 150\,\text{MeV}$, and some of them form a bound state 
with nuclei by the electromagnetic interaction. 
In next section, we consider non-standard nuclear reactions in the bound states, 
and find a solution to the $^{7}$Li problem.

\section{Non-standard nuclear reactions in stau-nucleus bound state}  
\label{Sec:non-standard} 

The long-lived staus form bound states with various nuclei, and the bound 
states open non-standard nuclear reactions. 
The internal conversion process is one of the non-standard nuclear reaction, 
which is analogous to electron capture reaction: 
$(N\,\tilde{\tau}^{-}) \to \tilde{\chi}_{1}^{0} + \nu_{\tau} + N'$. Here 
$(N \,\tilde{\tau}^{-})$ represents a bound state, and $N (N')$ is a nucleus. 
The internal conversion processes sufficiently destruct $^{7}$Be and $^{7}$Li, 
and lead to a solution to the $^{7}$Li problem (Sec.~\ref{Sec:Internal}). 
Another non-standard nuclear reaction is spallation processes: 
$(N \,\tilde{\tau}^{-}) \to \tilde{\chi}_{1}^{0} + \nu_{\tau} + N' + N'' + ...$\,. 
The spallation processes of $^4$He could enhance the production of neutron (n), deuteron (D), 
and triton (T), that leads to a strict bound on properties of the long-lived stau 
(Sec.~\ref{Sec:He4}).

\subsection{Internal conversion}  \label{Sec:Internal}  

\begin{figure}[ht]
\begin{center}
\includegraphics[width=127mm]{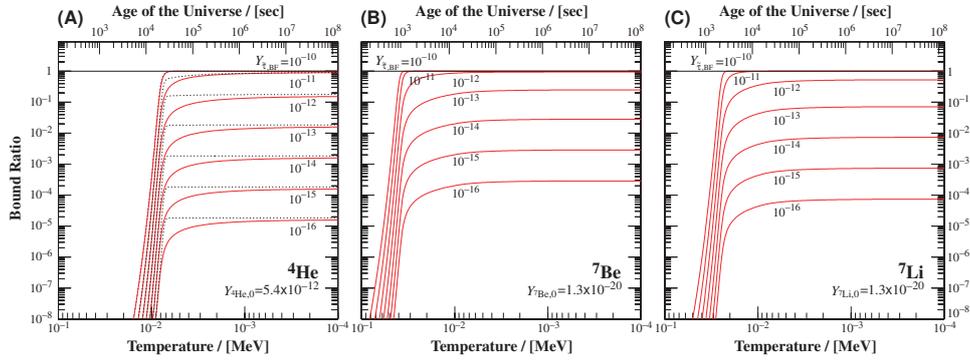}
\caption{
Evolutions of the bound ratio of the nuclei $\mathrm{^{4}He}$,
$\mathrm{^{7}Be}$, and $\mathrm{^{7}Li}$~\cite{Jittoh:2008eq}. 
We vary the abundance of the stau at the time of the formation of 
the bound state from $10^{-10}$ to $10^{-16}$ in each figure.
In Fig.\,\ref{Fig:bound-ratio}(A), we also plotted by dotted lines 
corresponding curves predicted using the Saha equation for reference.}
\label{Fig:bound-ratio}  
\end{center}
\end{figure}

\begin{figure}[ht]
\begin{center}
\includegraphics[width=70mm]{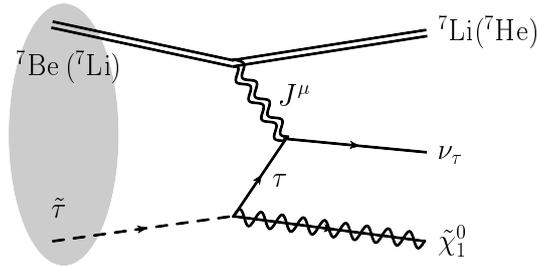}
\caption{The Feynmann diagrams of internal conversion of 
${}^7\text{Be}$ (${}^7\text{Li}$).}
\label{Fig:internal_diagrams}
\end{center}
\end{figure}

\begin{figure}[t]
\begin{center}
\begin{tabular}{cc}
\includegraphics[width=62mm]{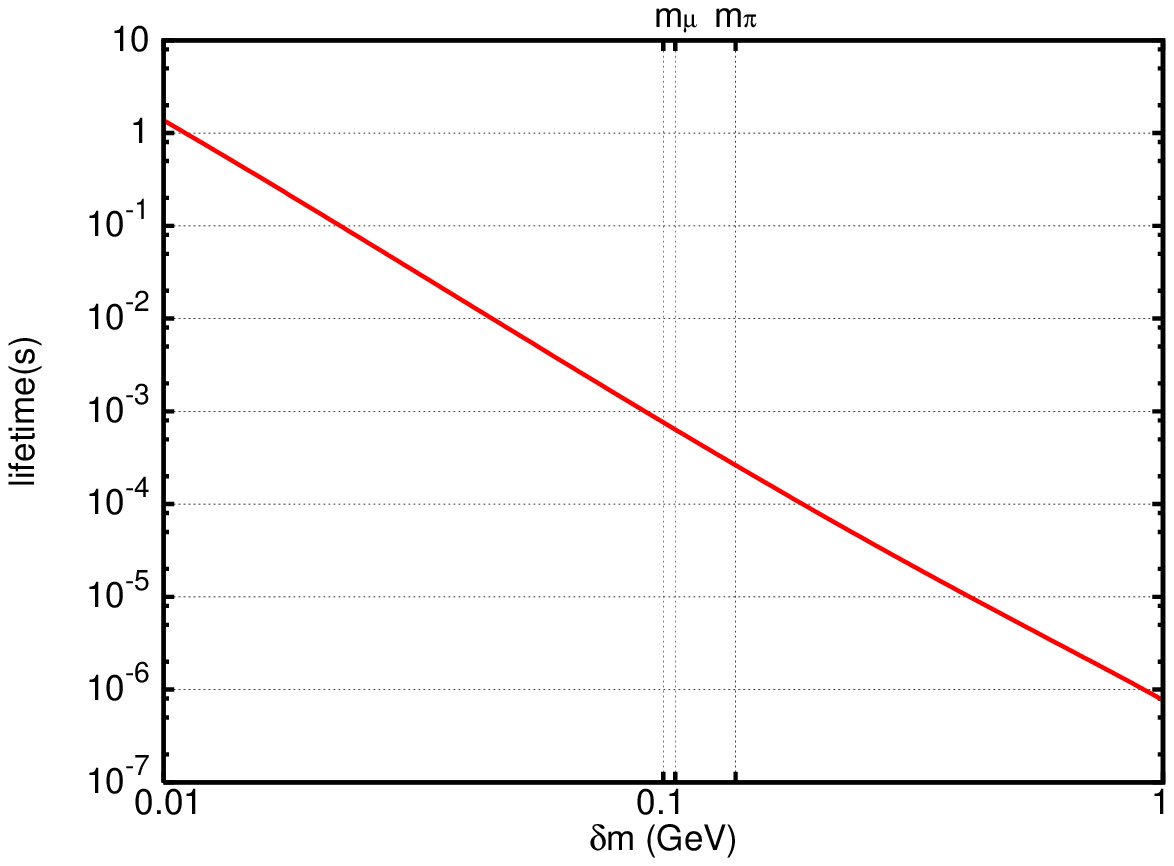} &
\includegraphics[width=62mm]{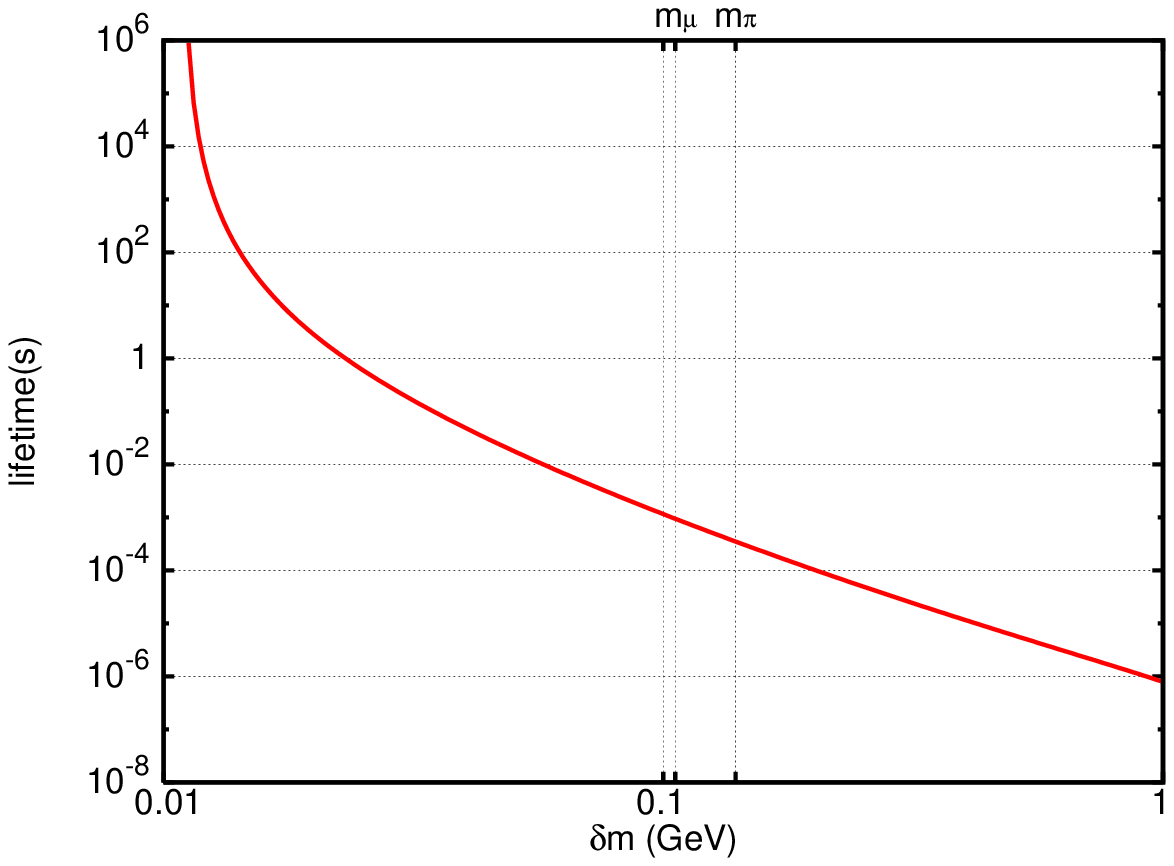}
\end{tabular}
\end{center}
\caption{
The lifetimes of internal conversion processes as the function of $\delta m$~\cite{Jittoh:2007fr}. 
Left panel: $({}^7\text{Be} \,\tilde{\tau}^{-}) \to \tilde{\chi}_{1}^{0} 
+ \nu_{\tau} + {}^7\text{Li}$, right panel: $({}^7\text{Li} \,\tilde{\tau}^{-}) 
\to \tilde{\chi}_{1}^{0} + \nu_{\tau} + {}^7\text{He}$. 
We take $m_{\tilde{\tau}}$ = 300GeV, $\theta_{\tau}=\pi/3$, 
and $\gamma_{\tau}=0$ for both reactions. 
}
\label{Fig:bound_lifetime}
\end{figure}

In a relatively early stage of the BBN, the long-lived stau forms a bound state with 
$^{7}$Be and $^{7}$Li nucleus (Fig.~\ref{Fig:bound-ratio}). These bound 
states give rise to the internal conversion processes 
(Fig.~\ref{Fig:internal_diagrams})~\cite{Jittoh:2007fr, Bird:2007ge}, 
\begin{subequations}
\begin{align}
    (^{7}\text{Be}\,\tilde{\tau}^{-}) 
    &\to 
	\tilde{\chi}_{1}^{0} + \nu_{\tau} + {}^{7}\text{Li},
    \label{Eq:internal_Be}
    \\
    (^{7}\text{Li}\,\tilde{\tau}^{-}) 
    &\to 
	\tilde{\chi}_{1}^{0} + \nu_{\tau} + {}^{7}\text{He}. 
    \label{Eq:internal_Li}
\end{align}
\label{Eq:internal_BeLi}
\end{subequations}
\\[-4mm]
The daughter $^{7}$Li nucleus in the process \eqref{Eq:internal_Be} 
is destructed either by an energetic proton or the process \eqref{Eq:internal_Li}. 
The daughter $^{7}$He nucleus in the process \eqref{Eq:internal_Li} 
immediately decays into a $^{6}$He nucleus and a neutron. 
One may consider that $^6$Li nucleus is produced via beta-decay of this $^6$He, and 
therefore the internal conversion processes cause the overproduction of $^6$Li.\footnote{This 
$^6$Li production process is different process from the catalyzed fusion of 
$^4$He$+$D that Pospelov pointed out in Ref.~\cite{Pospelov:2006sc}.} 
However, in the parameter region 
of our interest, $^7$Li is mainly destructed by a background proton into harmless nuclei.
As a result, the bound state formation of $^7$Li and stau is negligible, 
and hence we can safely ignore the $^6$Li production 
by the process of \eqref{Eq:internal_Li}.
Hence the non-standard chain reactions by the long-lived stau could yield 
smaller $^{7}$Be and $^{7}$Li abundances than those in the standard BBN 
scenario, that is the requirement for solving the $^{7}$Li problem (see 
Introduction). This is the scenario we proposed.

The reaction rate of the internal conversion processes is given by 
\begin{equation}
\begin{split}
	\Gamma_\text{IC}
	= 
	|\psi|^2 \cdot (\sigma v)_\text{IC}. 
	\label{Eq:ICrate}
\end{split}
\end{equation}
The front part in the right-hand side expresses overlap of wave functions in an initial state. 
We  assume that the bound state is in the S state of a hydrogen-like atom, 
and obtain the overlap as $|\psi|^2 = 1/(\pi a_{\text{nucl}}^{3})$. Here 
$a_{\text{nucl}} = (1.2 \times A^{1/3}) \, \mathrm{fm}$ is the radius 
of a nucleus and $A$ is its mass number. 
The remaining part in the right-hand side expresses the cross section of 
elementary process. For the process $({}^{7}\text{Be} \,\tilde{\tau}^{-}) 
\to \tilde{\chi}_{1}^{0} + \nu_{\tau} + {}^{7}\text{Li}$, 
\begin{equation}
\begin{split}
	&(\sigma v)_{\text{IC}}
	= 
	\frac{1}{(2E_{\tilde \tau}) (2E_\text{Be})} \int
	\frac{d^3 \boldsymbol{p}_\nu}{(2\pi)^3 2E_\nu} 
	\frac{d^3 \boldsymbol{p}_{\tilde \chi}}{(2\pi)^3 2E_{\tilde \chi}} 
	\frac{d^3 \boldsymbol{p}_{\text{Li}}}{(2\pi)^3 2E_{\text{Li}}}
	\\ & \hspace{2mm} \times
	\bigl| \mathcal{M} \bigl( ({}^{7}\text{Be} \,\tilde{\tau}^{-}) \to 
	\tilde{\chi}_{1}^{0} \nu_{\tau} {}^{7}\text{Li} \bigr) \bigr|^{2} 
	(2 \pi)^4 \delta^{(4)} 
	(p_{\tilde{\tau}} + p_\text{Be} - p_{\tilde \chi} - p_{\nu} - p_\text{Li}). 
\end{split}     \label{Eq:cross_IC}
\end{equation}
Here $p_i$ and $E_i$ are the momentum and the energy of a particle species $i$, 
respectively. The amplitude is decomposed into leptonic and hadronic part as 
\begin{equation}
\mathcal{M} 
\left( 
({}^{7}\text{Be} \,\tilde{\tau}^{-}) \to 
\tilde{\chi}_{1}^{0} \nu_{\tau} {}^{7}\text{Li} 
\right)
= 
\langle 
{}^{7}\text{Li} |J^{\mu}| {}^{7}\text{Be} 
\rangle  \, 
\langle 
\tilde{\chi}_{1}^{0} \, \nu_{\tau} |j_{\mu}| \tilde{\tau} 
\rangle.
\end{equation} 
The leptonic part is straightforwardly calculated. 
The matrix element of the nuclear conversion is evaluated by the $ft$ value 
obtained from $\beta$ decay experiments. The experimental $ft$ value is 
available for $^{7}\text{Li} \leftrightarrow {}^{7}\text{Be}$ but not 
for $^{7}\text{Li} \leftrightarrow {}^{7}\text{He}$. 
We assume that the two processes \eqref{Eq:internal_BeLi} have the same 
$ft$ value.  As long as we 
consider the quantum numbers of the ground state of ${}^{7}\text{Li}$ and 
${}^{7}\text{He}$, it is reasonable to expect a Gamow-Teller transition can occur 
since they are similar with those of ${}^{6}\text{He}$ and ${}^{6}\text{Li}$ 
and we know that they make a Gamow-Teller transition.  The Gamow-Teller 
transition is superallowed and has a similar $ft$ value to the Fermi transition 
such as $^{7}\text{Li} \leftrightarrow {}^{7}\text{Be}$.

The time scale of the internal conversion processes, \eqref{Eq:internal_Be} 
and \eqref{Eq:internal_Li}, are shown in Fig.\,\ref{Fig:bound_lifetime} 
as a function of $\delta m$. 
The time scale of $({}^{7}\text{Li} \,\tilde{\tau}^{-}) \to 
\tilde{\chi}_{1}^{0} + \nu_{\tau} + {}^{7}\text{He}$ diverges around 
$\delta m = m_{^{7}\text{He}} - m_{^{7}\text{Li}} = 11.7 \, \text{MeV}$, 
below which the internal conversion is kinematically forbidden.

We find that the reaction time scale is in the order of $10^{-3} \, \text{sec}$. 
Thus, compared with a typical time scale at the BBN epoch, a parent nucleus is 
converted into another nucleus in no time once the bound state is formed. 
The bound state formation makes the interaction between the stau and a 
nucleus more efficiently by two reasons: 
(i) the overlap of the wave functions of the stau and a nucleus becomes large since 
these are confined in the small space, 
(i\hspace{-1pt}i) the short distance between the stau and a nucleus 
allows virtual exchange of the hadronic current even if $\delta m < m_{\pi}$.

It is important to emphasize that the internal conversion processes can be activated 
only in the neutralino LSP scenarios, but not in the gravitino LSP scenarios. 
A key difference between these scenarios is the origin of longevity of the stau. 
In the neutralino LSP scenarios, the longevity is ensured by tight phase 
space suppression in its decay. Dominant channels of the particle decay of the 
stau are 3- and 4-body final state processes. The internal conversion processes 
are also 3-body final state processes since the processes occur via off-shell 
hadronic current exchange in the bound state, but the initial states are 2-body system, 
i.e., the stau and a nucleus. 
Consequently, compared with the particle decay of the stau, the phase space 
suppression is relaxed, and hence the time scale becomes much shorter than 
the stau lifetime. 
On the other hand, in the gravitino LSP scenarios, the longevity is ensured by 
the Planck suppressed interaction in its decay. Dominant decay channel is the 
2-body one, $\tilde{\tau} \to \tilde{G} \tau$. In this scenario, even in the 
bound state, the 2-body decay process dominates over the internal conversion 
processes, because the internal conversion needs the hadronic current which 
is suppressed by the Fermi couplings. Thus the internal conversion processes 
in the gravitino LSP scenario can not be activated.

\subsection{Spallation reactions and stau catalyzed fusion}  \label{Sec:He4} 

\begin{figure}[t]
\begin{center}
  \includegraphics[width=127mm,clip]{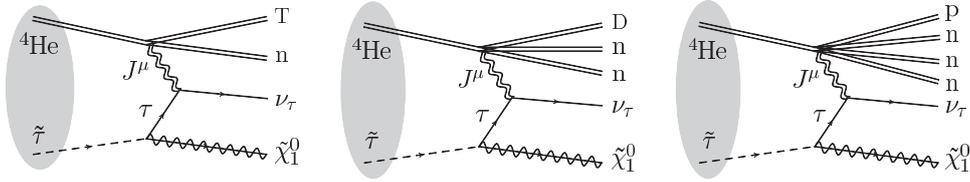}
  \caption{$^4$He spallation processes.}
  \label{Fig:spa_diagram}
  \end{center}
\end{figure}

There exists two types of non-standard reactions for the bound state of a stau and a 
$^4$He nucleus: 
\\
(1) stau-catalyzed fusion 
\begin{equation}
	(\mathrm{^{4}He} \,\tilde{\tau}^{-}) + \mathrm{D}
	\to
	\tilde{\tau} + \mathrm{^{6}Li}, 
	\label{Eq:catalyzed}
\end{equation}
(2) spallation of the $\mathrm{^{4}He}$ nucleus 
\begin{subequations}
\begin{align}
    ({}^{4}\text{He} \,\tilde{\tau}^{-}) 
    &\to 
	\tilde{\chi}_{1}^{0} + \nu_{\tau} + \text{T} + \text{n},
    \label{eq:spal-tn}
    \\
    ({}^{4}\text{He} \,\tilde{\tau}^{-}) 
    &\to 
	\tilde{\chi}_{1}^{0} + \nu_{\tau} + \text{D} + \text{n} + \text{n}, 
    \label{eq:spal-dnn}
	\\ 
	({}^{4}\text{He} \,\tilde{\tau}^{-}) 
	&\to 
	\tilde{\chi}_{1}^{0} + \nu_{\tau} + \text{p} + \text{n} + \text{n} + \text{n}. 
	\label{eq:spal-pnnn}
\end{align}
	\label{Eq:spa}
\end{subequations}
\\[-4mm]
Since the origin of longevity of the stau determines which light element is 
over-produced by these non-standard reactions, we need to carefully and 
independently study the evolution of the stau-${}^{4}\text{He}$ bound 
state in each scenario that possesses the origins of the longevity different from each 
other.

The catalyzed fusion process enhances the $^6$Li production~\cite{Pospelov:2006sc}. 
The catalyzed fusion process is a type of photon-less nuclear transfer reaction, 
and typical rate of such a reaction is formulated by using the astrophysical S-factor as 
$S(0) \sim 10^{2}\,\text{keV\,b}$~\cite{Angulo:1999zz}. 
Thermal averaged cross section of the catalyzed fusion is precisely calculated in 
Refs.~\cite{Hamaguchi:2007mp, Kamimura:2008fx}, which is much larger than that 
of the $^6$Li production in the SBBN, $^{4}\text{He} + \text{D} \to {}^{6}\text{Li} 
+ \gamma$,  by 6-7 orders of magnitude. 
In the SBBN, since the $^6$Li production $^{4}\text{He} + \text{D} \to 
{}^{6}\text{Li} + \gamma$ is an E1 forbidden transition reaction, the reaction rate is 
tiny and the magnitude is given by the astrophysical S-factor as $S(0) \sim 2 \times 
10^{-6}\,\text{keV\,b}$~\cite{Angulo:1999zz}. 
The over-production of $^{6}$Li nucleus by the catalyzed fusion process leads to 
stringent constraints both on the lifetime and the number density of the long-lived 
charged particles in the BBN epoch.~\cite{Pospelov:2006sc}.

Reaction \eqref{Eq:spa} is essentially a spallation of the $\mathrm{^{4}He}$ nucleus, 
producing a triton, a deuteron, and neutrons. We formulate the reaction rate of a 
spallation process $({}^{4}\text{He} \,\tilde{\tau}^{-}) \to \tilde{\chi}_{1}^{0} 
+\nu_{\tau}+\text{T}+\text{n}$. The reaction rate is given as 
\begin{equation}
\begin{split}
	\Gamma
	\left( 
	({}^{4}\text{He} \,\tilde{\tau}^{-}) \to 
	\tilde{\chi}_{1}^{0} + \nu_{\tau} + \text{T} + \text{n}
	\right)
	= 
	|\psi|^2 \cdot \sigma v_{\textrm{Tn}}. 
	\label{time_Tn}
\end{split}
\end{equation}
With the same formulation in the case of internal conversion process, the overlap 
of wave functions of the stau and the $\mathrm{^4He}$ nucleus is calculated 
as $|\psi|^2 = (Z \alpha m_\text{He})^3/\pi$ where $Z$ is the atomic number 
of nucleus and $\alpha$ is the fine structure constant. 
The cross section $\sigma v_{\textrm{Tn}} \equiv \sigma v 
\left( ({}^4\text{He} \,\tilde{\tau}^{-}) \to \tilde{\chi}_{1}^{0} 
\nu_\tau \text{Tn} \right)$ 
is calculated as follows, 
\begin{equation}
\begin{split}
	\sigma v_{\textrm{Tn}}
	&= 
	\frac{1}{2E_{\tilde \tau}} \int
	\frac{d^3 \boldsymbol{p}_\nu}{(2\pi)^3 2E_\nu} 
	\frac{d^3 \boldsymbol{p}_{\tilde \chi}}{(2\pi)^3 2E_{\tilde \chi}} 
	\frac{d^3 \boldsymbol{q}_\text{n}}{(2\pi)^3}
	\frac{d^3 \boldsymbol{q}_\text{T}}{(2\pi)^3}  
	\bigl| \mathcal{M} \bigl( ({}^4\text{He} \,\tilde{\tau}^{-}) \to 
	\tilde \chi_1^0 \nu_\tau \text{Tn} \bigr) \bigr|^2 
	\\ & \hspace{2mm} \times
	(2 \pi)^4 \delta^{(4)} 
	(p_{\tilde \tau} + p_\text{He} - p_\nu - q_\text{T} - q_\text{n}). 
\end{split}     \label{cross_tn_1}
\end{equation}

\begin{figure}[t]
\begin{center}
\includegraphics[width=95mm]{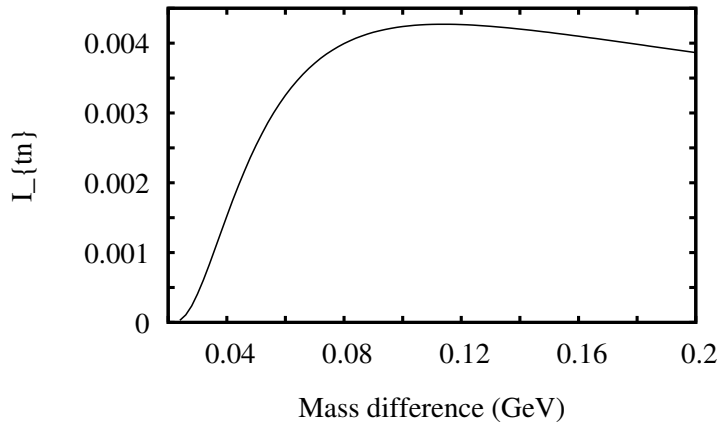}
\caption{Numerical result of $I_\text{Tn}$ as a function of 
$\delta m = m_{\tilde{\tau}} - m_{\tilde{\chi}_{1}^{0}}$ 
for $m_{\tilde \tau} = 350$GeV. }
\label{Fig:Itn}
\end{center}
\end{figure}

The amplitude is decomposed into leptonic and hadronic part as 
\begin{equation}
\mathcal{M} 
\left( 
({^4}\text{He} \,\tilde{\tau}^{-}) \to \tilde{\chi}_{1}^{0} 
\nu_{\tau} \text{Tn} 
\right) 
= 
\langle 
\text{Tn} |J^{\mu}| {^4}\text{He} \rangle  \, 
\langle \tilde{\chi}_{1}^{0} \hspace{0.5mm} 
\nu_{\tau} |j_{\mu}| \tilde{\tau} 
\rangle. 
\end{equation}
The hadronic current $J^{\mu}$ is given by a vector current $V_{\mu}$ 
and an axial vector current $A_{\mu}$ as $J_{\mu} = V_{\mu} + 
g_\text{A} A_{\mu}$, where $g_\text{A}$ is the axial coupling constant.
The relevant components are $V^{0}$ and $A^{i}$ ($i = 1, 2, 3$). 
We take these operators as a sum of single-nucleon operators as
\begin{equation}
  V^{0}
  =
  \sum_{a = 1}^{4} \tau_{a}^{-}
  \mathrm{e}^{\mathrm{i}\boldsymbol{q} \cdot \boldsymbol{r}_{a}}
  \, , \quad
  A^{i}
  =
  \sum_{a = 1}^{4} \tau_{a}^{-} \sigma_{a}^{i}
  \mathrm{e}^{\mathrm{i}\boldsymbol{q} \cdot \boldsymbol{r}_{a}}
  \, ,
\end{equation}
where $\boldsymbol{q}$ is the momentum carried by the current, 
$\boldsymbol{r}_{a}$ are the spatial coordinates of the $a$-th nucleon ($a \in 
\{1, 2, 3, 4 \}$), and $\tau_a^-$ and $\sigma_{a}^{i}$ denote the isospin 
ladder operators and the spin operators of the $a$-th nucleon, respectively. 
Each component leads to a part of hadronic matrix element:
\begin{equation}
\begin{split}
	\langle \text{Tn} \left| V^{0} \right| \hspace{-0.5mm} ^{4}\text{He} \rangle 
	= 
	\langle \text{Tn} \left| A^{+} \right| \hspace{-0.5mm} ^{4}\text{He} \rangle
	= 
	- \langle \text{Tn} \left| A^{-} \right| \hspace{-0.5mm} ^{4}\text{He} \rangle
	= 
	- \langle \text{Tn} \left| A^{3} \right| \hspace{-0.5mm} ^{4}\text{He} \rangle
	= 
	\sqrt{2} \mathcal{M}_{\textrm{Tn}}, 
	\label{Eq:had_mat} 
\end{split}
\end{equation}
where $A^\pm = (A^1 \pm i A^2)/\sqrt{2}$. 
We need wave functions of $^{4}$He, T, and n to calculate the hadronic amplitude. 
The formulation and explicit formulae of the wave functions are given in Appendix A 
in Ref.~\cite{Jittoh:2011ni}.
Consequently we obtain the hadronic matrix element as follows, 
\begin{equation}
\begin{split}
   &
   \mathcal{M}_{\textrm{Tn}} 
   = 
   \biggl(
   \frac{128 \pi}{3} 
   \frac{a_\text{He} a_\text{T}^2}{(a_\text{He} + a_\text{T})^4}
   \biggr)^{3/4} 
   \\ &~~ \times
   \biggl\{ 
   \exp \biggl[ - \frac{\boldsymbol{q}_\text{T}^2}{3 a_\text{He}} \biggr] 
   - \exp \biggl[ - \frac{\boldsymbol{q}_\text{n}^2}{3 a_\text{He}} 
   - \frac{(\boldsymbol{q}_\text{T} + \boldsymbol{q}_\text{n})^2}
   {6 (a_\text{He} + a_\text{T})} \biggr]
   \biggr\}.
   \label{amp_part} 
\end{split}
\end{equation}
\begin{table}
\caption{Input values of the matter radius $R_\text{mat}$ 
 for D, T, and $^4\text{He}$, the magnetic radius 
$R_\text{mag}$ for p and n,  nucleus mass $m_X$, excess 
energy $\Delta_X$ for the nucleus $X$, and each reference. }
{\begin{tabular}{llll} \hline
nucleus  
& $R_\text{mat(mag)}$ [fm]/[GeV$^{-1}$] ~
& $m_X$ [GeV] 
& 
$\Delta_X$ [GeV] 
\\[0.7mm] \hline
p
& 0.876 / 4.439 \cite{Borisyuk:2009mg}
& 0.9383 \ \cite{Mohr:2008fa}
& $6.778 \times 10^{-3}$ \ \cite{TOI}
\\[0.7mm] \hline
n
& 0.873 / 4.424 \cite{Kubon:2001rj}
& 0.9396 \ \cite{Mohr:2008fa}
& $8.071 \times 10^{-3}$ \ \cite{TOI}
\\[0.7mm] \hline
D          
& 1.966 / 9.962 \ \cite{Wong:1994sy}
& 1.876 \ \cite{TOI}  ~
& $1.314 \times 10 ^{-2}$ \ \cite{TOI}
\\[0.7mm] \hline
T           
& 1.928 / 9.770 \ \cite{Yoshitake}  
& 2.809 \ \cite{TOI}  
& $1.495 \times 10 ^{-2}$ \ \cite{TOI}
\\[0.7mm] \hline
$^4$He 
& 1.49 / 7.55 \ \cite{Egelhof2001307}
& 3.728  \ \cite{TOI}
& $2.425 \times 10 ^{-3}$ \ \cite{TOI}
\\[0.7mm] \hline
\end{tabular} \label{table:input}}
\end{table}
Here $\boldsymbol{q}_\text{T}$ and $\boldsymbol{q}_\text{n}$ are three-momenta
of the triton and the neutron, respectively, and $a_\text{He}$ and $a_\text{T}$
are related to the mean square matter radius $R_\text{mat}$ by 
$a_\text{He} = 9/(16 (R _{\text{mat}})_{\text{He}}^{2})$ and 
$a_\text{T} = 1/(2 (R _{\text{mat}})_{\text{T}}^{2})$, respectively. 
We list in Table\,\ref{table:input} input values of the matter radius. 
The leptonic part is straightforwardly calculated to be 
\begin{equation}
\begin{split}
	\left| \langle 
	\tilde{\chi}_{1}^{0} \hspace{0mm} \nu_{\tau} \left| j_{0} \right| \tilde{\tau} 
	\rangle \right|^{2}
	&= 
	\left| \langle 
	\tilde{\chi}_{1}^{0} \nu_{\tau} \left| j_{z} \right| \tilde{\tau} 
	\rangle \right|^{2} 
	= 
	4 G_\text{F}^{2} \left| g_\text{R} \right|^{2} 
	\frac{m_{\tilde{\chi}_{1}^{0}} E_{\nu}}{m_{\tau}^{2}},
	\\ 
	\left| \langle 
	\tilde{\chi}_{1}^{0} \nu_{\tau} \left| j_{\pm} \right| \tilde{\tau} 
	\rangle \right|^2 
	&= 
	4 G_\text{F}^{2} \left| g_\text{R} \right|^{2} 
	\frac{m_{\tilde{\chi}_{1}^{0}} E_{\nu}}{m_{\tau}^{2}} 
	\biggl( 
	1 \mp \frac{p_{\nu}^{z}}{E_{\nu}}
	\biggr) , 
	\label{leptonic_part}
\end{split} 
\end{equation}
where $E_\nu$ and $p_\nu^z$ are the energy and the $z$-component of the 
momentum of the tau neutrino, respectively. Consequently we obtain the cross 
section as follows, 
\begin{equation}
\begin{split}
   \sigma v_\text{Tn} 
   &= 
   \frac{8}{\pi^2} \biggl( \frac{32}{3 \pi} \biggr)^{3/2}
   g^2 \tan^2\theta_W \sin^2\theta_\tau (1+3g_A^2) G_F^2 
   \Delta_\text{Tn}^4 
   \ \frac{m_\text{T} m_\text{n}}{m_{\tilde \tau} m_\tau^2} \ 
   \frac{a_\text{He}^{3/2} a_\text{T}^3}{(a_\text{He} + a_\text{T})^5} 
   ~I_\text{Tn} , 
   \label{cross_tn_2} 
\end{split} 
\end{equation}
where $I_\text{Tn}$ is a dimensionless integral, which includes kinematical 
information of the reaction. The numerical result of $I_\text{Tn}$ is plotted in 
Fig.~\ref{Fig:Itn}, and the analytical formula is given by Eq.~(17) in 
Ref.~\cite{Jittoh:2011ni}. 
Here $\Delta_\text{Tn}$, $k_\text{T}$, and $k_\text{n}$ are defined as 
$\Delta_\text{Tn} \equiv \delta m + \Delta_\text{He} - \Delta_\text{T} - 
\Delta_\text{n} - E_\text{b}$, 
$k_\text{T} \equiv \sqrt{2 m _{\text{T}} \Delta_\text{Tn}}$, and 
$k_\text{n} \equiv \sqrt{2 m _{\text{n}} \Delta_\text{Tn}}$. 
Here $\Delta_X$ is the excess energy of the nucleus $X$, and $E_\text{b}$ 
is the binding energy of $({}^4\text{He} \,\tilde{\tau}^{-})$ system.

We compare the rate of the spallation and that of the stau-catalyzed fusion. 
We first note that the rate of stau-catalyzed fusion strongly depends on the
temperature~\cite{Hamaguchi:2007mp}, and we fix the reference 
temperature to be $30\,\text{keV}$.
Staus begin to form a bound state with $^4$He at this temperature, which 
corresponds to cosmic time of $10^{3}\,\text{s}$. Thus the bound 
state is formed when the lifetime of the staus is longer than $10^{3}\,\text{s}$.

\begin{figure}[t]
\begin{center}
\includegraphics[width=95mm]{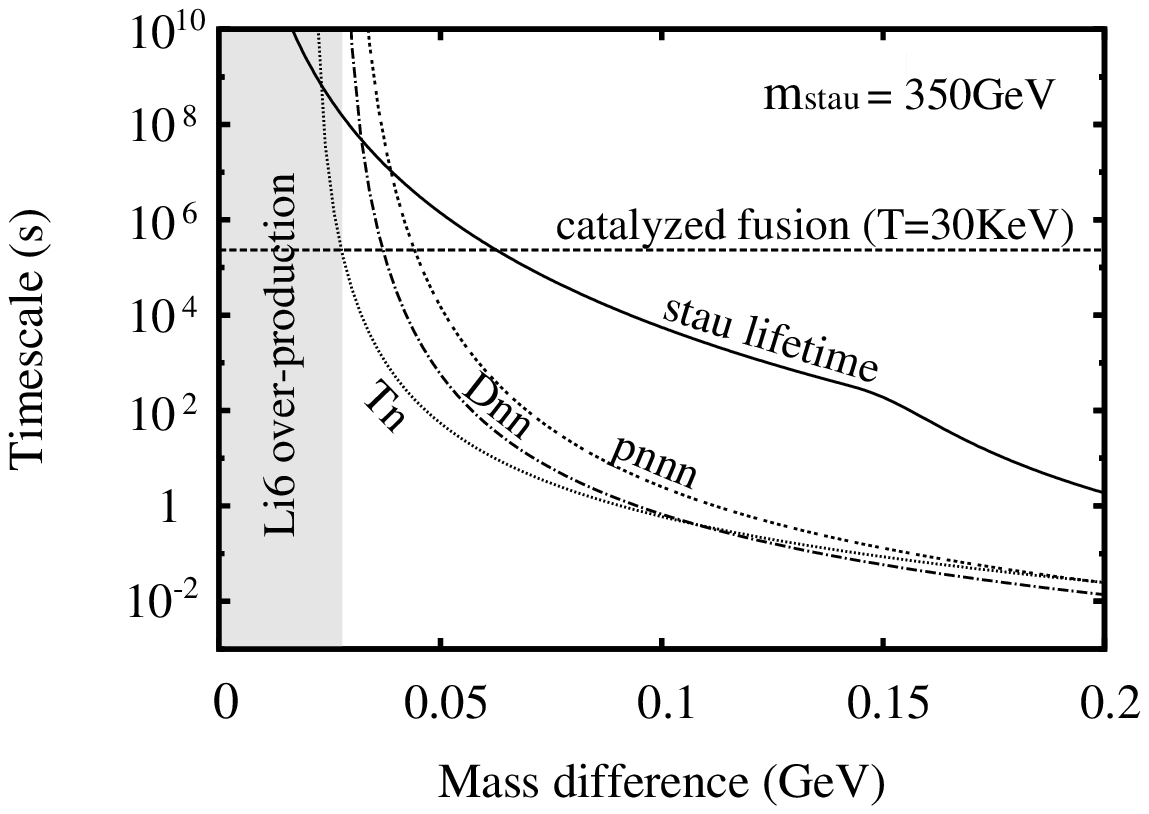}
\caption{Timescale of spallation processes as a function 
of $\delta m$~\cite{Jittoh:2011ni} 
and the stau-catalyzed fusion at the 
universe temperature $T=30$\,keV~\cite{Hamaguchi:2007mp}. 
The lifetime of free $\tilde{\tau}$ (solid line) is also depicted.  
Here we took $m_{\tilde \tau} = 350$GeV, $\sin\theta_\tau 
= 0.8$, and $\gamma_\tau = 0$.}
\label{fig:time}
\end{center}
\end{figure}

Figure~\ref{fig:time} shows the timescale of the spallation processes as a 
function of $\delta m$. The lifetime of free stau is plotted by a solid line. 
We took the reference values of $m_{\tilde \tau} = 350$\,GeV, 
$\sin\theta_\tau = 0.8$, and $\gamma_\tau = 0$. The timescale of 
the stau-catalyzed fusion at the temperature of $30\,\text{keV}$ is also 
shown by the horizontal dashed line.
Once a bound state is formed, as long as the phase space of spallation 
processes are open sufficiently that is $\delta m \gtrsim 0.026$\,GeV, 
those processes dominate over other processes.  There $\tilde \tau$ 
property is constrained to evade the over-production of D and/or T. 
For $\delta m \lesssim 0.026$\,GeV, the dominant process of 
$({}^4\text{He} \,\tilde{\tau}^{-})$ is stau-catalyzed fusion, since the 
free $\tilde \tau$ lifetime is longer than the timescale of stau-catalyzed 
fusion. Thus light gray region is forbidden due to the over-production 
of $^6$Li.

The result of Fig.\ref{fig:time} is not much altered by varying the parameters 
relevant with $\tilde \tau$.  First the cross sections of the spallation processes are 
inversely proportional to $m_{\tilde \tau}$,  and then the timescale of each 
process linearly increases as $m_{\tilde \tau}$ increases. Thus, even when 
$m_{\tilde \tau}$ is larger than $m_{\tilde \tau} = 350$\,GeV by up to a 
factor of ten, the region of $^6$Li over-production scarcely changes.  
Next we point out that our result depends only mildly on the left-right mixing 
of the stau.  Indeed, the cross section of the $^4 \text{He}$ spallation is 
proportional to $\sin ^2 \theta _{\tau}$.  The order of its magnitude will not 
change as long as the right-handed component is significant.

\section{Numerical Results}

In this section, we show numerical results of the number density and the 
allowed parameter region of the stau by solving the Boltzmann equations at 
BBN era. Details of numerical calculations are explained in 
Ref.~\cite{Jittoh:2008eq}.

\subsection{Stau Number Density}

\begin{figure} [t]
\begin{center}
\begin{tabular}{cc}
\includegraphics[width=180pt, clip]{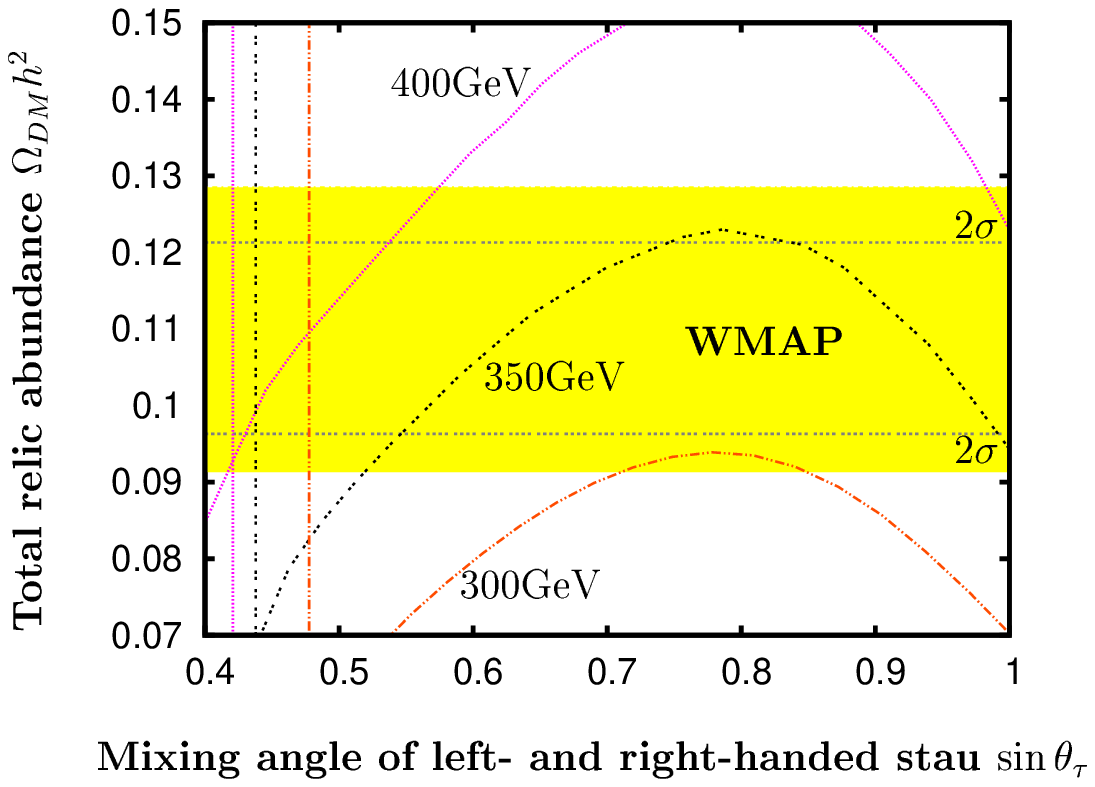} &
\includegraphics[width=180pt, clip]{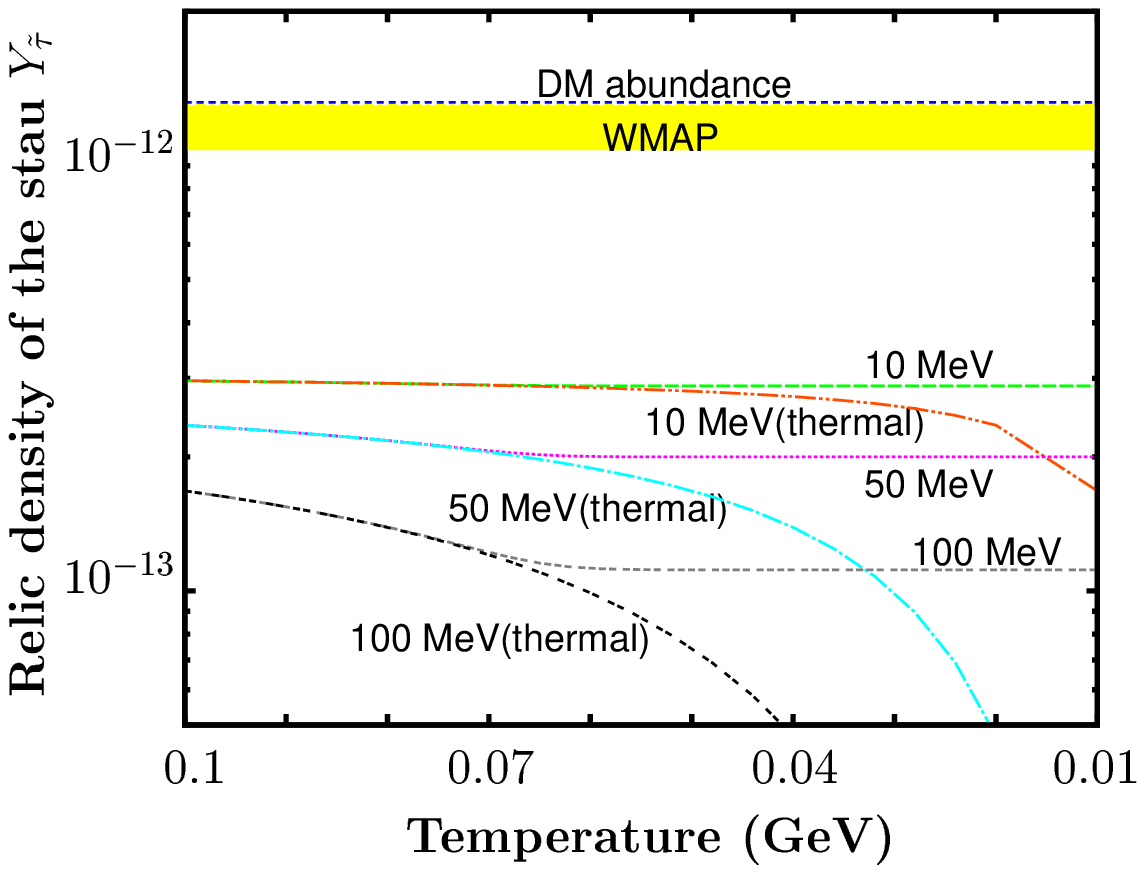}
\end{tabular}
\caption{Left: Total abundance of staus and neutralinos~\cite{Jittoh:2010wh}.
Right: The evolution of the number density of negative charged stau~\cite{Jittoh:2010wh}.
}
\label{total}
\end{center}
\end{figure} 

The neutralino LSP is the dark matter in our scenario, hence its abundance 
must coincide with the observed dark matter abundance at the present time. 
Most of the neutralino LSP is produced via decays of the long-lived stau during 
the BBN. Thus, before the BBN starts, the total abundance of stau and the 
neutralino LSP must be the same as that of the dark matter.

We  show the total abundance of the staus and the neutralino LSP in terms of left-right mixing parameter. 
In the left panel of Fig.~\ref{total}, 
each curve corresponds to the stau mass $300,~350$ and $400$ GeV, respectively, and horizontal (yellow) band and the dotted lines represent 
the allowed region from the WMAP observation at the 3$\sigma$ level ($0.0913 \leq \Omega_{DM} h^2 \leq 0.1285$) and 
at the $2\sigma$ level  ($0.0963 \leq \Omega_{DM} h^2 \leq 0.1213$) \cite{Dunkley:2008ie}.
The left-side region from the vertical lines in the figure, corresponding to $m_{\tilde \tau} = 400, 350, 300 \,\mathrm{GeV}$ from left to right, 
represents the left-handed sneutrino LSP. 
The left-handed sneutrino dark matter has been ruled out by constraints from 
the direct detection experiments \cite{Falk:1994es} and therefore
only the right-side region is allowed. Here we took $\gamma_\tau = 0$ and 
$\delta m = 100 \, \mathrm{MeV}$.

We can see that the total abundance increases first as the heavier stau mixes to the lighter
stau, then turns to decrease at $\sin\theta_{\tau} \simeq 0.8$.  The increase
of the abundance can be understood by the fact that the annihilation cross section 
of $\tilde{\tau} + \tilde{\tau} \rightarrow \tau + \tau$,  $\tilde{\tau}^\ast + \tilde{\tau}^\ast \rightarrow \bar{\tau} + \bar{\tau}$ and $\tilde{\tau} + \tilde{\tau}^\ast \rightarrow \tau + \bar{\tau}$
becomes smaller as the
heavier stau mixes. The increase is gradually compensated by two annihilation
processes, $\tilde{\tau} + {\tilde{\tau}}^\ast \rightarrow W^+ + W^-$ and 
$\tilde{\chi}^0_1 + \tilde{\tau} \rightarrow W^- + \nu_\tau$, as the left-right 
mixing becomes large. Here $\tilde{\tau}^\ast$ denotes anti-particle of stau.
The latter process can not be ignored because left-handed 
sneutrino is degenerate to stau and neutralino in the present parameter set. These processes 
become significant for $\sin\theta_{\tau} < 0.8$. Another process which reduces 
the total abundance due to the left-right mixing is $\tilde{\tau} + 
\tilde{\tau}^\ast \rightarrow t + \bar{t}$ through s-channel exchange of the 
heavy Higgses. This annihilation process becomes significant as the mixing reaches 
to $\pi/4$ and the masses of two staus are split. In the left panel, the mass difference 
between the lighter and the heavier stau is fixed to be $30$ GeV to maximize the 
dark matter abundance, but the same result can be obtained by changing the stau 
mass for another values of the mass difference.
As is seen, the total abundance also strongly depends on $m_{\tilde \tau}$.
This is understood as follows.
In the non-relativistic limit, since the relic number density of relic species is 
proportional to $(m_{relic} \langle \sigma v \rangle_{\textrm{sum}})^{-1}$ 
and $\langle \sigma v \rangle_{\textrm{sum}}$ is proportional 
to $1/m_{relic}^2$~\cite{Kolb}, the total number density $N$ is proportional to 
$m_{\tilde \tau}$,
\begin{equation}
\begin{split}
    N \propto \frac{1}{m_{\tilde \tau} \langle \sigma v \rangle_{\textrm{sum}}} 
    \propto \frac{1}{1/m_{\tilde \tau}}
    = m_{\tilde \tau} ~,
 \end{split}
 \label{k}
\end{equation}
and the total abundance is given by $\Omega_{DM} h^2 \sim m_{\tilde \tau} N$.
Thus the total abundance is proportional  to $m_{\tilde \tau}^2$, and it is
consistent with the result in Fig. \ref{total}.

The right panel of Fig.~\ref{total} shows the evolution of the number density of stau as a
function of the universe temperature. Here we took $m_{\tilde \tau} = $
350 GeV, $\sin \theta_\tau$ = 0.8, and $\gamma_\tau $ = 0 and chose
$\delta m$ = 10 MeV, 50 MeV, and 100 MeV as sample points.
Each line with attached  [$\delta m$] shows the number density of stau, while the 
one with attached  [$\delta m$(thermal)] shows the equilibrium one.
Horizontal dotted line represents the relic density of dark matter, which is the
total abundance and used as an initial condition of total value for the number density ratio. Yellow band represents the
allowed region from the WMAP observation at the $2\sigma$ level \cite{Dunkley:2008ie}.

The number density evolution of stau is qualitatively understood as follows. As
shown in the figure, the freeze-out temperature of stau is almost independent of $\delta m$.
It is determined by the exchange processes, $\tilde{\tau} \tau^{+} \leftrightarrow 
\tilde{\chi}_{1}^{0} \gamma$ and $\tilde{\tau} \gamma \leftrightarrow 
\tilde{\chi}_{1}^{0} \tau^{-}$, whose magnitude $\langle \sigma v \rangle 
Y_{\tilde \tau} Y_{\gamma}$ is governed by the factor $e^{-(m_\tau - \delta m)/T}$, 
where $m_\tau$ represents the tau lepton mass.
The freeze-out temperature of the stau density $T_{f \textrm{(ratio)}}$ is given
by $(m_{\tau} - \delta m)/T_{f \textrm{(ratio)}} \simeq 25$,
since the cross section of the exchange process is of the same magnitude as weak
processes.
Thus $T_{f \textrm{(ratio)}}$ hardly depends on $\delta m$.
In contrast, the ratio of the number density between stau and neutralino depends
on $\delta m$, $n_{\tilde \tau} / n_{\tilde \chi}
\sim \text{exp} (- \delta m / T)$, since they follow the Boltzmann distribution
before their freeze-out.
Thus, the relic density of stau strongly depends on $\delta m$.

Here, we comment on the dependence of the stau relic density $n_{\tilde \tau^-}$
on other parameters such as $m_{\tilde \tau}$, $\theta_{\tau}$, and
$\gamma_{\tau}$. 
The number density of the negatively charged stau is expressed in terms of the total
relic density $N$ by
\begin{equation}
 \begin{split}
   n_{\tilde \tau^-} 
   =
   \frac{N}{2 (1 + e^{\delta m/T_{f \textrm{(ratio)}}})} ~.
 \end{split}     \label{l}
\end{equation} 
Here, the freeze-out temperature $T_{f \textrm{(ratio)}}$ hardly depends on
these parameters.
This is because the cross section of the exchange processes are changed by these
parameters at most by factors but not by orders of magnitudes, and the $T_{f
  \textrm{(ratio)}}$ depends logarithmically on 
$\langle \sigma v \rangle$\cite{Kolb, Jittoh:2010wh}.
On the other hand, the total relic density $N$ is proportional to $m_{\tilde
  \tau}$ as in Eq.~(\ref{k}).
The value of $N$ is also affected by the left-right mixing $\theta_{\tau}$ as
seen in Fig.~\ref{total} since the annihilation cross section depends on this
parameter.
In contrast, $\gamma_{\tau}$ scarcely affects the relic density, since this
parameter appears in the annihilation cross section through the cross terms of the
contributions from the left-handed stau and the right-handed one, and such terms
are always accompanied by the suppression factor of $m_{\tau}/m_{\tilde{\tau}}$
compared to the leading contribution.
Thus the relic number density of stau $n_{\tilde{\tau}}$ strongly depends on
$m_{\tilde{\tau}}$ and $\theta_{\tau}$ while scarcely depends on
$\gamma_{\tau}$.

After the number density of stau freezes out, stau decays according to its
lifetime \cite{Jittoh:2005pq}, or forms a bound state with a nucleus in the BBN
era.
Their formation rate has been studied in literatures~\cite{Kohri:2006cn,
  Jittoh:2007fr, Bird:2007ge}.
The bound states modify the predictions of SBBN, and make it possible to solve
the $^7$Li problem via internal conversion processes in the bound state
\cite{Jittoh:2007fr, Bird:2007ge, Jittoh:2008eq}.

\begin{figure}[t!]
\begin{center}
\begin{tabular}{cc}
    \includegraphics[width=80mm]{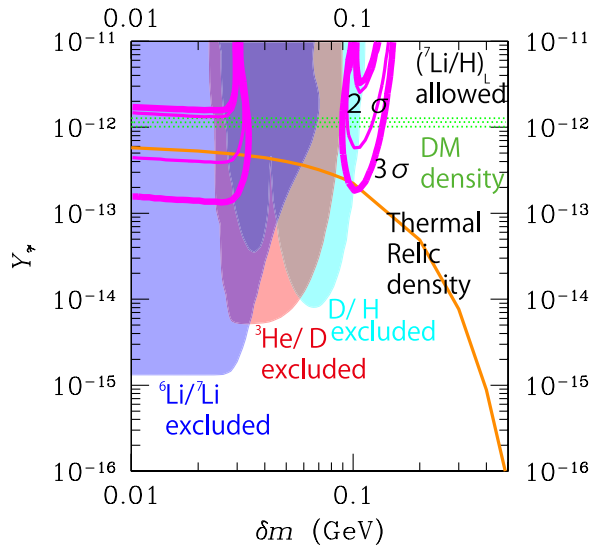} 
\end{tabular}
    \caption{
    Allowed regions from observational light element abundances at 2 $\sigma$~\cite{Jittoh:2011ni}.  
    The higher value of the observational  $^7$Li/H in \cite{Melendez:2004ni} denoted by
    $(^7$Li/H)$_{\rm H}$ are adopted. 
    Thin and thick lines represent the 2$\sigma$ and 3$\sigma$ only for $^7$Li/H, respectively. 
    The horizontal band means the observationally-allowed dark matter density. 
    The parameters are taken as $m_{\tilde{\tau}} = 350$~GeV, $\sin\theta_{\tau} = 0.8$, and $\gamma_{\tau}=0$.  
    }
    \label{fig:y_deltamMR}
 \end{center}
\end{figure}

%
\subsection{Parameter region for the solution of $^7$Li Problem}

Next, we show the allowed regions of the parameter space in which the 
$^7$Li problem can be solved.  We include in our analyses not only the internal conversion 
processes but also $^4$He spallation processes and $\tilde \tau$ catalyzed nuclear fusion  explained 
in the previous section.

As mentioned in the introduction, the $^7$Li problem is a discrepancy between the
theoretical value of $^7$Li abundance predicted in the SBBN and the observational one. 
The prediction of the SBBN for the $^7$Li to H ratio is $^7$Li/H $=4.47^{+0.58}_{-0.66} \times 10^{-10}$
for baryon to photon ratio $\eta = (6.225 \pm 0.170) \times 10^{-10}$ 
(68\% C.L.) by the WMAP \cite{Komatsu:2010fb}. 
On the other hand, the primordial $^7$Li abundance is observed in
metal-poor halo stars as absorption lines~\cite{Spite:1982dd}. Observationally-inferred values of the 
primordial $^7$Li to H ratio is 
$^7$Li/H $=2.34 ^{+0.30}_{-0.35} \times 10^{-10}$
for a high value\cite{Melendez:2004ni}. 
(See also Refs.~\cite{Ryan:2000zz, Asplund:2005yt,  Aoki:2009ce, Bonifacio:2006au} for another values.)
Therefore the discrepancy remains more than $3\sigma$ 
between theoretical and observational values.

In Fig.~\ref{fig:y_deltamMR}, we plot the allowed parameter region in $\delta m$ and $Y_{\tilde{\tau}}$ plane, 
which is obtained by comparing the theoretical values to observational ones for the high value of $^7$Li/H.
The parameters for the stau are taken to be $m_{\tilde \tau} = $ 350 GeV, $\sin \theta_\tau$ = 0.8, and
$\gamma_\tau$ = 0.
We have adopted
following another observational constraints on the light element
abundances: an upper bound on the $^6$Li to $^7$Li ratio,
$^6$Li/$^7$Li $< 0.046 + 0.022$~\cite{Asplund:2005yt}, the deuteron to
hydrogen ratio, D/H=$(2.80 \pm 0.20) \times
10^{-5}$~\cite{Pettini:2008mq}, and an upper bound on the $^3$He to
D ratio, $^3$He/D $< 0.87 + 0.27$~\cite{GG03}.
The solid (orange) line denotes a theoretical value of the  thermal relic
abundance for the stau~\cite{Jittoh:2010wh} while keeping
observed dark matter density at $\Omega_{\rm DM}h^{2} =
0.11 \pm 0.01$ (2$\sigma$)~\cite{Komatsu:2010fb}  as  total
$\tilde{\chi}_{1}^{0}+\tilde{\tau}$ abundance. For reference, we also plot the
observationally-allowed dark matter density in the figures by a
horizontal band.

At around $\delta m \sim 0.1$ GeV, we found that there is an allowed region at $Y_{\tilde{\tau}} \sim
10^{-13}$ for $\delta m \sim 130$~MeV to solve the $^7$Li problem at $3\sigma$.
It is shown that $^7$Li/H can be fitted to the observational value without conflicting with the other light element
abundances.~\footnote{See also~\cite{Jedamzik:2004er, Bird:2007ge, Pospelov:2010cw, Kawasaki:2010yh, Erken:2011vv} 
for another mechanisms to reduce
$^7$Li/H. } As shown in Fig.~\ref{fig:y_deltamMR},  it is impressive that the relic density is consistent with the  allowed
region at $Y_{\tilde{\tau}} = 2 \times 10^{-13}$ at 3 $\sigma$ in case
of the high value of $^7$Li/H.
On the other hand, since the stau lifetime is longer than the formation timescale of the 
stau-${}^{4}\text{He}$ bound state in the region $\delta m \lesssim 100\,\text{MeV}$, 
over-productions of $\text{D}$ and ${}^{3}\text{He}$ by the ${}^{4}\text{He}$ 
spallation processes exclude the region $30\,\text{MeV} \lesssim \delta m \lesssim 
100\,\text{MeV}$ and $Y_{\tilde{\tau}} \gtrsim 10^{-14}$. 
In the smaller $\delta m$ region, the ${}^{4}\text{He}$ spallation processes 
kinematically close, and the catalyzed fusion is activated. Consequently, 
the observational ${}^{6}\text{Li}/{}^{7}\text{Li}$ ratio excludes 
the region where $Y_{\tilde{\tau}^{-}} \gtrsim 10^{-15}$ and $\delta m 
\lesssim$100\,MeV. 
This feature is explained as follows.

We have adopted following observational abundances of  $^6$Li and
$^7$Li.
Throughout this subsection, $n_{i}$ denotes the number density of a particle
``$i$'', and observational errors are given at $1 \sigma$.
For the $n_{\rm ^6Li}$ to $n_{\rm ^7Li}$ ratio, we use the upper
bound~\cite{Asplund:2005yt},
\begin{eqnarray}
    \label{eq:Li6obs}
    \left( n_{\rm ^6Li}/n_{\rm ^{7}Li} \right)_p
     < 0.046 \pm 0.022 + 0.106,
\end{eqnarray}
with a conservative systematic error (+0.106)~\cite{Hisano:2009rc}, 
while the $^7$Li abundance is given above.
In the current scenario $^6$Li can be overproduced by the catalyzed fusion \eqref{Eq:catalyzed}.
The abundance of $^{6}$Li through this process is
approximately represented by
\begin{eqnarray}
    \label{eq:DeltaYLi6}
    \Delta Y_{^{6}{\rm Li}} \sim \frac{\langle \sigma v
    \rangle_{^{6}{\rm Li}}n_{\rm
    D}}{H} Y_{\tilde{\tau}^{-}},
\end{eqnarray}
with $\langle \sigma v \rangle_{^{6}{\rm Li}}$ the thermal average of the cross
section times the relative velocity for this process~\cite{Hamaguchi:2007mp},
and $n_{\rm D}$ the number density of deuterium.
Note that in Eq.\eqref{eq:DeltaYLi6}, it is roughly assumed that all staus have been
bound to $^4$He when the $^6$Li production proceeds via the reaction \eqref{Eq:catalyzed}.
By using (\ref{eq:Li6obs}), we see that the additional $^6$Li production is
constrained to be $\Delta Y_{^{6}{\rm Li}} < {\cal O}(10^{-21})$.
Numerical value of $\langle \sigma v \rangle_{^{6}{\rm Li}}$ gives $\langle
\sigma v \rangle_{^{6}{\rm Li}} n_{\rm D}/ H \sim {\cal O}(10^{-6})$ at $T\sim
$10\,keV.
Then from (\ref{eq:DeltaYLi6}) it is found that the upper bound on the abundance
of stau should be $Y_{\tilde{\tau}^{-}} \lesssim 10^{-15}$.
Because the bound state $(^{4}{\rm He} \, \tilde{\tau}^{-})$ forms at $T \lesssim
$10 keV, this process is strongly constrained for $\tau_{\tilde{\tau}^{-}}
\gtrsim 10^{4}$s with $\tau_{\tilde{\tau}^{-}}$ being the stau lifetime, which
corresponds to $\delta m \lesssim 100$ MeV.
Note that the ratio $\langle \sigma v \rangle_{^{6}{\rm Li}} n_{\rm D}/ H $
rapidly decreases as the cosmic temperature decreases, and this late 
production of $^{6}$Li is much more effective just after formation of the bound state.
This is the reason why we can estimate (\ref{eq:DeltaYLi6}) at around
10\,keV.

On the other hand, the rates of $^{7}$Be and $^{7}$Li destruction through the
internal conversion~\cite{Jittoh:2007fr,Bird:2007ge,Jittoh:2008eq} could
be nearly equal to the formation rates of the bound state 
$({}^{7}{\rm Be} \, \tilde{\tau}^{-})$ and $({}^{7}{\rm Li} \, 
\tilde{\tau}^{-})$, respectively.
This is because the timescale of the destruction through the internal conversion
is much faster than that of any other nuclear reaction rates and the Hubble
expansion rate.
Then the amount of destroyed $^{7}$Be (or $^{7}$Li after its electron capture)
is approximately represented by
\begin{eqnarray}
    \label{eq:Delta7}
   \Delta Y_{^{7}{\rm Be}} \sim  \frac{\langle \sigma v \rangle_{\rm
    bnd,7}n_{^{7}{\rm Be}}}{H} Y_{\tilde{\tau}^{-}},
\end{eqnarray}
where $\langle \sigma v \rangle_{\rm bnd,7} \sim 10^{-2} {\rm
GeV}^{-2} (T/30{\rm keV})^{-1/2}(Z/4)^{2} \times(A/7)^{-3/2}
(E_{b^{7}{\rm Be}}/1350 {\rm keV})$ is the thermally-averaged cross
section times the relative velocity of the bound-state formation for
$(^{7}{\rm Be}\tilde{\tau}^{-})$ ~\cite{Kohri:2006cn,Bird:2007ge}. 
Eq.\eqref{eq:Delta7} gives an approximate order of magnitude only under the condition of 
$Y_{\mathrm{stau}} < Y_{^4\mathrm{He}}$.
We require $\Delta Y_{{}^{7}{\rm Be}}$ to become $\sim {\cal O}(10^{-20})$
to reduce the abundance of ${}^{7}{\rm Be}$ to fit the observational
data. Then the abundance of $\tilde{\tau}^{-}$ should be the order of
$ \Delta Y_{{}^{7}{\rm Be}} (\langle \sigma v \rangle_{\rm bnd,7}
n_{{}^{7}{\rm Be}}/H )^{-1} \sim {\cal O}(10^{-12})$ with $\langle
\sigma v \rangle_{\rm bnd,7} n_{{}^{7}{\rm Be}}/H \sim 10^{-8}$ at $T$ =
30 keV. Because $\langle \sigma v \rangle_{\rm bnd,7} n_{^{7}{\rm
Be}}/H$ decreases as the cosmic temperature decreases ($\propto
T^{1/2}$), the destruction is more effective just after the formation of
the bound state. This validates that we have estimated
(\ref{eq:Delta7}) at 30 keV. Therefore the parameter region at around
$Y_{\tilde{\tau}^{-}} \sim 10^{-12}$ and $\delta m \lesssim $~130~MeV
is allowed by the observational $^{7}$Li. Here $\delta m \lesssim
$~130~MeV corresponds to $\tau_{\tilde{\tau}} \gtrsim 10^{3}$~s. The
case for the destruction of (${}^{7}$Li~$\tilde{\tau}^{-}$) through the
internal conversion is also similar to that of
(${}^{7}$Be~$\tilde{\tau}^{-}$)~\cite{Jittoh:2007fr,Jittoh:2008eq}.

Further constraints come from the relic density of the dark matter, which 
can be stated in terms of the stau relic density.
It is calculated as shown in Fig.~\ref{fig:y_deltamMR} for the present values
of parameters.
Applying all the constraints, we are led to the allowed interval shown by the
thick line in the figure.

\subsection{Constraint on parameter space of stau}
\label{LHC} 

\begin{figure} [t!]
\begin{center}
\includegraphics[width=250pt,clip]{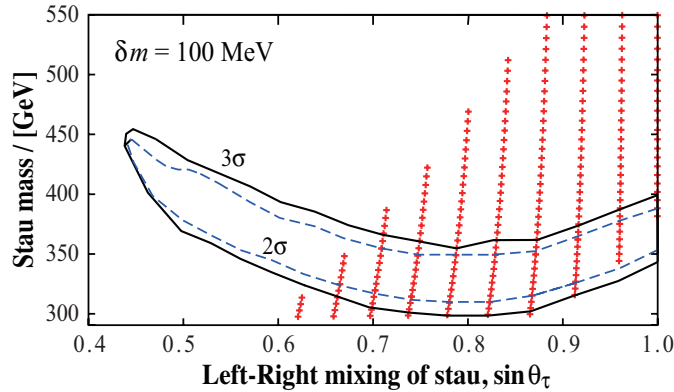}
\caption{ %
  Parameter space of the stau for the observational abundances of the 
  dark matter and that of the light elements~\cite{Jittoh:2010wh}.
  The region surrounded by black solid (blue dashed) line shows
  the allowed region from the WMAP observation at the 3$\sigma$
  (2$\sigma$) level \cite{Dunkley:2008ie}.  Red crisscross points show
  the parameters which is consistent with observational abundances for the
  light elements including $^7$Li at $3\sigma$ level. } \label{allowed}
\end{center}
\end{figure}

In the end, we show in Fig.~\ref{allowed} the parameter space in which the
dark matter abundance and light elements abundances are consistent with 
the observations.
Here, based on the discussion in previous subsection, we took $\delta m =
100 \,\mathrm{MeV}$.
Parameter region surrounded by black solid (blue dashed) line is allowed
by the relic abundance of the dark matter from the WMAP observation at the $3\sigma$
($2\sigma$) level \cite{Dunkley:2008ie}, which corresponds to $0.0913
\leq \Omega_{\textrm{DM}} h^2 \leq 0.1285$ ($0.0963 \leq
\Omega_{\textrm{DM}} h^2 \leq 0.1213$).
Red crisscross points show the parameters which are consistent with the 
observational abundances for the light elements including $\mathrm{^7Li}$, 
where the observational $\mathrm{^7Li}$ abundance is given in 
Ref.~\cite{Melendez:2004ni}.

The abundance of the light elements constrains the parameter space due to the
following reasons.
First, the region where the stau mass is less than 300\,GeV is excluded since the
relic density becomes too small to destruct $\mathrm{^{7}Li}$ sufficiently. 
Next, the top-left region of the figure is excluded since the lifetime of
stau becomes too long and hence overproduces $\mathrm{^6Li}$ through the
catalyzed fusion \cite{Pospelov:2006sc}. 
The stau lifetime becomes longer as its mass becomes heavier~\cite{Jittoh:2005pq}.
On the other hand, the lifetime becomes shorter as the left-right mixing angle
increases in the present parameter space~\cite{Jittoh:2005pq}.
As a result, the allowed region is obtained as shown by the red crisscrosses
in Fig.~\ref{allowed}.

The black solid curve and the blue dashed curve are the constraints from the relic
abundance of the dark matter as discussed in the previous section.
Note that the relic abundance is not so sensitive to the mass difference $\delta m$ 
for $\delta m \ll m_{\tilde \chi}$.
Combination of the constraints on the abundance of the light elements
and of the dark matter strongly restricts the allowed region and leads to
$\delta m \simeq 100 \,\mathrm{MeV}$, $m_{\tilde \tau} = (300 \textrm{
-- } 400) \,\mathrm{GeV}$, and $\sin\theta_{\tau} = (0.65 \textrm{ -- }
1)$.
In Fig.~\ref{fig:y_deltamMR}, these parameter values correspond to
the region which is surrounded by the curves of the thermal relic density (orange) 
and $^7$Li/H with 
$3\sigma$ significance (purple).
Our model can thus provide a nice probe to the mixing angle, which has few
experimental signals, once the value of $m_{\tilde \tau}$ is determined.

\section{Summary}

In this review we expounded our idea to fit the primordial abundance of 
$^7$Li to the observed value. We extend the SM to the supersymmetric SM 
(SSM). To explain the bino-like neutralino dark matter abundance with the 
coannihilation mechanism in the SSM, we have a long-lived scalar particle, stau.
It can be so long-lived that it affects the BBN prediction.
With appropriate parameters we can nicely fit the abundances both of $^7$Li 
and the neutralino dark matter to the observed ones.

The key property is the small mass difference between the stau and the 
neutralino dark matter. It gives the following important features: 
\begin{itemize}
\item[(1)]
Very long life time of $\mathcal{O}$(1000)\,s since the phase 
space is quite narrow and the open decay channel is only the 4 
body decay (in Fig.~\ref{Fig:diagrams} (c))
\item[(2)]
The stau in (${}^{7}\text{Be} \,\tilde{\tau}^{-}$) very efficiently
destructs $^7$Be, because the phase space becomes much wider in 
the bound state than that of the free stau decay. 
This is shown in Fig.~\ref{Fig:internal_diagrams}. Comparing it with the
Fig.~\ref{Fig:diagrams} (b), it is understood as the opening of the threshold
of the 3 body decay. Therefore the time scale of the destruction is of order 
1\,s or less.
\end{itemize}
These two features are available only in the case where the stau longevity is 
ensured by small mass difference, not in the case where the origin of the 
longevity is Planck suppressed coupling.

With such a stau we have to consider the following three processes in 
addition to the standard processes in the SBBN: 
\begin{itemize}
\item[(a)]
The destruction process of $^7$Be (and $^7$Li) in the bound state with 
the stau (Sec.~3.1). It is often called ``internal conversion''. This is 
important to reduce the primordial $^7$Li abundance. More reactions 
make the prediction better. 
\item[(b)]
The $^4$He spallation processes in the bound state 
$({}^{4}\text{He} \, \tilde{\tau})$ (Sec.~3.2). These processes give 
extra light elements such as D, $^3$He. Less reactions are better. 
To evade the over-productions of light elements, the stau must not be 
long-lived so that it decays before (${}^{4}\text{He} \,\tilde{\tau}^{-}$) 
is formed or $\delta m$ must be large enough as shown in Fig.~\ref{fig:time}. 
\item[(c)]
The $^6$Li production process called catalyzed fusion (\eqref{Eq:catalyzed} 
in Sec.~3.2). If we take the $^6$Li problem seriously, appropriate  reaction 
rate of the catalyzed fusion would be needed.
>From this requirement we can pin down the parameter space if necessary.
\end{itemize}

We also evaluate the stau number density at the BBN era which is important 
and necessary to solve the Boltzmann equations of nucleosynthesis chain. 
Larger stau density makes the $^7$Li abundance lower, and the larger 
stau density is obtained for smaller mass difference of the stau and the neutralino.

Again we emphasize that the first two processes (a) and (b) are specific to 
our scenario, or more exactly to the small mass difference scenario.
On the contrary the third process (c) can happen in all scenarios with 
long-lived charged massive particles.

As long as tau flavor is conserved exactly all the rate for the processes 
(a)-(c) and the stau number density are the function of only the mass 
difference $\delta m$ eventually. In other words the prediction depends 
almost only on the mass difference.

Including all the processes we calculated the primordial abundance
and we showed that we possibly have the solution to the $^7$Li problem
under the assumption of lepton flavor conservation.

The upper bound on $\delta m$ is mainly determined by the requirement
that the stau be long-lived enough to be relevant with the nucleosynthesis 
chain reactions. The lower bound also determined by the requirement that
the stau be not too long-lived so that it would not destruct $^4$He so much.
Thus, in turn, to solve the ${}^{7}\text{Li}$ problem, we have very 
narrow allowed region for the parameter space. It gives a very precise 
prediction for particle physics like a collider phenomenology.

Before concluding the review we give a comment  on the
case that there is lepton flavor violation~\cite{Kohri:2012gc}.
Including this we can relax the constraint on $\delta m$ and we 
have a room to produce $^6$Li appropriately  by the catalyzed fusion.
In this cases, for example, stau mixes with scalar electron,
\begin{eqnarray}
 \tilde{\tau} \rightarrow 
 \tilde{l} = \tilde{\tau}+c_e\tilde{e},
\end{eqnarray}
where $\tilde{l}$ is a slepton, and $c_e$ denotes a tiny mixing parameter 
with the selectron. The slepton can decay into a neutralino and an electron, 
\begin{eqnarray}
 \tilde{l} \rightarrow \tilde{\chi}^{0}_{1}+e .
\end{eqnarray}
Therefore the lifetime of the slepton will depend on not only $\delta m$ 
but also $c_e$. The mixing parameter $c_e$ must be very small, of 
$\mathcal{O}$($10^{-10}$), so that the stau still be long-lived. With 
this degree of freedom we have a wider room for $\delta m$. 
The over-production of D and ${}^{3}\text{He}$ by the $^4$He 
spallation processes (b) can be suppressed since $\delta m$ can be very 
small. Therefore we would not have any constraint from this process
as long as it is small enough.

On the contrary due to the smallness of $c_e$, the efficiency of the 
internal conversion process (a) still depends only on $\delta m$. 
The rate of the process is still rapid even if $\delta m\sim 30$\,MeV
as shown in Fig.~\ref{Fig:bound_lifetime}.

\begin{figure}[ht]
\begin{center}
\includegraphics[width=100mm]{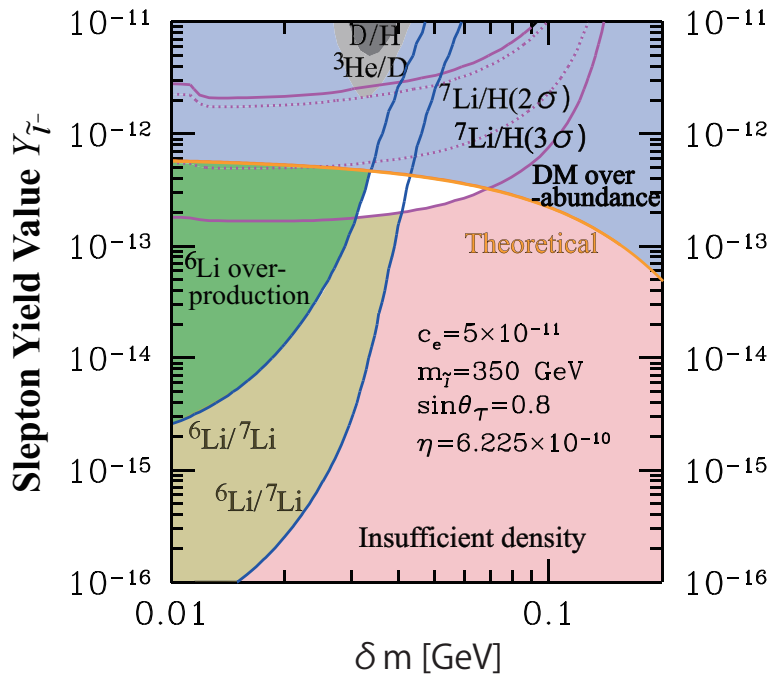}
\caption{
Allowed regions by observational light element abundances in
$\delta m$--$Y_{\tilde l^{-}}$ planes in cases of 
$c_{e}=5 \times 10^{-11}$~\cite{Kohri:2012gc}. 
The lines of the 
constraints are plotted for D/H (dark grey), $^{3}$He/D (light grey), and 
$^{6}$Li/$^{7}$Li (blue curves) at 2$\sigma$ . An exception is  $^7$Li/H (purple lines) 
whose constraints are denoted by both dotted lines at 2$\sigma$, 
and solid lines at 3$\sigma$. The theoretical curve of the slepton 
density is plotted as a thick solid (orange) line.}
\label{Fig:flavor_violation}
\end{center}
\end{figure}

By combining the smallness of $\delta m$ and $c_{e}$,  we get an 
allowed region (white) as shown in Fig.~\ref{Fig:flavor_violation} where 
$c_e= 5 \times 10^{-11}$.  Not only the $^7$Li problem is solved 
within 3 (and almost 2)$\sigma$ level, but also there is the ``allowed'' 
region for the $^6$Li problem. If we do not take the $^6$Li problem
seriously, from its abundance all the right region of $^{6}\text{Li}$ 
over-production is the allowed region for the $^6$Li abundance. 
If we take it seriously, 
the white band within two lines indicated $^6$Li/$^7$Li is the allowed
region and in near future once the $^6$Li abundance is measured 
definitely we can pin down the allowed region within the white band. 
Thus, in this case both the $^6$Li and the $^7$Li problems are solved.

To conclude our review, we emphasize that we did not introduce 
any artificial  particles. We employed the most extensively studied
physics beyond the SM, that is, SSM. Within this theory we showed that
the lithium problem(s) can be solved. We use the parameter which is
required not by the lithium problem but by the explanation for dark matter
abundance. In this sense  our solution is very excellent!

\section*{Acknowledgments}
This work is supported in part by the Grant-in-Aid for the Ministry of Education, 
Culture, Sports, Science, and Technology, Government of Japan, 
No.~25105009 (J.S.), No.~25003345 (Y.M.) and No.~15K17654 (T.S.).

\bibliographystyle{apsrev}
\bibliography{sample}
\end{document}